# Voltage gated inter-cation selective ion channels from graphene nanopores

*Lauren Cantley[1], Jacob L. Swett[2], David Lloyd[1], David A. Cullen[3], Ke Zhou[4], Peter V. Bedworth[2], Scott Heise[2], Adam J. Rondinone[3], Zhiping Xu[4], Steve Sinton[2], J. Scott Bunch[1,5]\**

[1]Department of Mechanical Engineering, Boston University, Boston, Massachusetts 02215, USA

[2]Advanced Technology Center, Lockheed Martin Space, Palo Alto, California, 94304, USA

[3]Center for Nanophase Materials Sciences, Oak Ridge National Laboratory, Oak Ridge, Tennessee, 37831, USA

[4]Applied Mechanics Laboratory, Department of Engineering Mechanics and Center for Nano and Micro Mechanics, Tsinghua University, Beijing 100084, China

[5]Division of Materials Science and Engineering, Boston University, Brookline, Massachusetts 02446, USA

*Correspondence to: bunch@bu.edu

**ABSTRACT. With the ability to selectively control ionic flux, biological protein ion channels perform a fundamental role in many physiological processes. For practical applications that require the functionality of a biological ion channel, graphene provides a promising solid-state alternative, due to its atomic thinness and mechanical strength. Here, we demonstrate that nanopores introduced into graphene membranes, as large as 50 nm in diameter, exhibit inter-cation selectivity with a ~20x preference for $K^+$ over divalent cations and can be modulated by an applied gate voltage. Liquid atomic force microscopy of the graphene devices reveals surface nanobubbles near the pore to be responsible for the observed selective behavior. Molecular dynamics simulations indicate that translocation of ions across the pore likely occurs via a thin water layer at the edge of the pore and the nanobubble. Our results**

**demonstrate a significant improvement in the inter-cation selectivity displayed by a solid-state nanopore device and by utilizing the pores in a de-wetted state, offers an approach to fabricating selective graphene membranes that does not rely on the fabrication of sub-nm pores.**

Protein ion channels, which are vital for many biological processes, including cell signaling and volume regulation within cells, are remarkably effective due to their high selectivity, permeability, and gating[1]. This has motivated the development of solid-state devices that mimic their function for practical applications in sensing, separation, therapeutics, and neuromorphic computing. Solid-state nanochannel and nanopore transistors have previously been used to manipulate ionic transport[2-3]; however, thus far they have been limited by low electrolyte concentrations[4-5], high applied voltages[3,6], or a combination of the two[7].

Graphene nanopores have been explored for applications in sensing and separations, and are a promising material for a solid-state ion channel. Aside from graphene being atomically thin[8], mechanically strong[9], and relatively inert[10], it has been shown that well-defined nanometer and sub-nanometer pores can be controllably introduced into the material[11-17]. Nanopores in graphene have been shown to exhibit ion selectivity[12, 18-24] and gated nanopores in graphene have been used in sensing biomolecules such as DNA and proteins[25-26]. However, graphene nanopores have yet to mimic the degree of inter-cation selectivity exhibited by protein ion channels.

For the nanopore devices studied here, single-layer graphene was obtained by CVD growth. Suspended graphene membranes were fabricated by transferring graphene over an approximately 5 µm diameter hole etched in a suspended silicon nitride window coated with 20 nm of atomic-layer-deposited (ALD) alumina. A gold electrode was patterned in contact with the suspended graphene membrane via a shadow mask. Suspended graphene devices were mounted in a custom-made microfluidic cell, allowing for the introduction of electrolyte solution to both sides of the graphene membrane. Measurement of conductance across the graphene membrane was carried out by applying a bias voltage across the device and measuring the resulting current (Fig.1a). The microfluidic cell allowed for electrical contact to the gold electrode, permitting the application of a gate voltage to the graphene while sweeping the transmembrane bias voltage. Additionally, the microfluidic cell was designed to allow atomic force microscopy (AFM) access to the top side of the membrane such that the pore could be imaged while in solution. Devices made from unperforated graphene had a trans-membrane conductance below 150 pS in 0.1 M KCl and 280 pS in 1 M KCl, confirming that graphene is a good barrier to ionic conductance. Graphene devices were perforated via helium ion microscope (HIM) drilling. Scanning transmission electron microscopy (STEM), HIM, and AFM were used to measure the pore diameters and observe the pore structure (Fig. 1b, 1c).

Ionic conductance was first measured in a two-terminal configuration across each device, with the gate terminal floating. For the graphene device in figure 1b, non-linear (activated) I-V characteristics were observed using monovalent electrolyte solutions (Fig. 2a). To account for the differences in bulk conductivity, the normalized conductance was plotted for each cation-chloride solution

$$g_i = \frac{G_i}{\sigma_i / \sigma_{KCl}} \qquad (1)$$

where $G_i$ is the measured nanopore conductance in solution $i$, $\sigma_i$ is the bulk conductivity of solution $i$ and $\sigma_{KCl}$ is the bulk conductivity of KCl at a comparable chloride concentration. The normalized conductance reveals the pore(s) to be highly cation selective, with significant-preference for $K^+$ over other ions measured. For the device shown in figure 1b at $V_s = 0$ mV, the normalized conductance of KCl was ~4x greater than the other monovalent ions ($Na^+$ and $Li^+$) and ~20x greater than the divalent ion measured ($Ca^{2+}$) (Fig. 2b). The differences in normalized conductance as well as the absence of conductivity in $CaCl_2$ suggest that the dominant charge carriers are cations; additional experiments using asymmetric ion conditions confirm this result (SI Appendix, Fig. S2).

Next, a gate voltage was applied to the graphene to modulate the ionic current. Before proceeding with voltage-gated measurements, leakage current from source/drain to gate was measured to be less than 300 pA at 500 mV, verifying the device conductance was governed by ion transport and not a result of leakage current. The pore current was then measured under various applied gate voltages. Figure 2c demonstrates the ionic current response to changes in the gate voltage. As a more negative gate voltage is applied, the ionic current increased. As positive gate voltage was applied, there was no significant change in the ionic conductivity. This unipolar behavior is similar to a p-type FET device, likewise suggesting that cations are the majority charge carriers, and is consistent with 2 terminal measurements[27-29].

To characterize the selectivity of a device, we define the conductance ratio as $S_i = g_i/g_{KCl}$. This definition gives a conductance ratio of 1 for a pore that does not distinguish between cation $i$ and $K^+$. The conductance ratio was measured for 10 graphene devices with HIM drilled pores. Six devices had nine approximately 30 nm diameter pores, the same configuration as the device shown in figure 1b, and four had a single 50 nm diameter pore, shown in figure 1c. Five of the six nine-pore devices and three of the four single pore devices displayed selective behavior. The mean conductance ratio for these devices is plotted in figure 2d. All samples displayed a similar trend in selectivity, where divalent ions had a lower conductance ratio than the monovalent species measured.

The activated IV behavior and observed selectivity in figure 2 inversely scales with the trend in hydrated radii of the measured cations, $K^+ < Na^+ < Li^+ < Ca^+ < Mg^+$ [1, 23], where $K^+$ has the highest conductance. Additionally, the ability to detect an electrostatic gating response in solution is dependent on the Debye screening length: a measure of a charge carrier's electrostatic range in solution (~1 nm and ~0.3 nm in 0.1 M and 1M KCl, respectively). The pore diameter of a fully wet pore should be within a given solution's Debye length in order to observe direct electrostatic gating effects. However, the discrepancy between imaged pore size and the observed selective gate-responsive behavior suggests that the pores are not fully wet. Similarly, the absolute value of the conductance across the graphene pore is lower than one would expect given a standard model for pore conductance based on the imaged pore diameter[30]; this also suggests incomplete wetting.

Incomplete wetting of a pore occurs often in nanopore experiments, particularly on hydrophobic surfaces[31]. Nanoscale surface bubbles are known to be present and highly stable on hydrophobic surfaces, such as highly oriented pyrolytic graphite, and occur in at least three types: gaseous nanobubbles, nanobubbles composed of oil[45], and solid nanoparticles[46]. They are often produced via the exchange of ethanol to aqueous solution, a procedure utilized in the wetting of our graphene devices[32]. STEM imaging of our devices reveals a concentration of hydrocarbons adsorbed onto the surface of the graphene near the pore (SI Appendix, Fig.S3). The presence of these surface adsorbates not only modifies the wettability and the surface charge of the pore, but provide favorable locations (such as step edges or defects) for a nanobubble to pin[33]. AFM imaging of a graphene device in water revealed a nanobubble on the surface of the graphene, occluding the pore (Figure 3a-c, SI Appendix, Fig.S4-S6). Subsequent conductance measurements across the device show selective activated I-V behavior similar to that observed in previous devices (Figure 3d, Figure 2). While the presence of nanobubbles was found to be ubiquitous, our limited control over the bubble formation may explain the variability in figure 2d. Within the devices studied, selective behavior was observed with the presence of a nanobubble occluding all or part of the pore area. Conversely, the device AFM imaged in water that did not possess a nanobubble displayed linear, non-selective I-V characteristics with a conductance consistent with the theoretically expected value for the imaged pore size (SI Appendix, Fig.S7-S8).

To better understand the ion translocation process across a nanobubble at the entrance of a graphene nanopore, we performed molecular dynamics (MD) simulations to explore the free energy profiles (SI Appendix, Fig.S1). Considering the limitation of computational cost, we simulated water-immersed porous graphene membranes containing nanopores with radii in the range of 0.9-2.0 nm, with graphene edges functionalized by carbonyl groups and a gaseous nanobubble partially occluding the pore. As illustrated in Figure 3e, two transport pathways were considered for the ion translocation. In path 1, ions travel through a thin 1 nm thickness water film coating the graphene edges, where the hydration shells (HSs) can be perturbed by the (functionalized) graphene edges. Here the water film thickness is defined as the distance between the water surface and the edge carbon atoms in graphene. In path 2, the HSs must be stripped off, for the ion to translocate across the water/gas interface.

We calculate the potential of mean force (PMF) from the MD simulation, to measure the change of free energy during the ion translocation process (Figure 4). For path 2, across the water/gas interface, a simple but over-estimation of the free energy barrier is $\Delta G = (1 - 1/\varepsilon)q^2/(4\pi\varepsilon_0 R) = \sim 10$ eV by using the Born model with the assumption that the HSs are fully detached from the ions[41]. Here $\varepsilon \approx 80$ is the relative permittivity of water and $\varepsilon_0$ is the vacuum permittivity. Our MD simulations show, however, that the HSs are partially retained as the ion translocates across the interface. The free energy barrier is reduced to 51.0 $k_BT$ as a result, which is still very significant compared to the thermal fluctuation, indicating that the hydrated ion prefers to stay in the solvent and transport along path 2 is prohibited (Figure 4a).

Conversely, the free energy barrier along path 1, through the thin water layer of 1 nm coating the graphene edge, is much reduced. Our free energy analysis shows that with the HSs perturbed, or ions captured by the functional groups at the graphene edge[44], the barriers for the adsorption-desorption process are on the order of $k_BT$. Specifically, the barrier for Na$^+$ with a perturbed 2$^{nd}$ or 1$^{st}$ HS is $\Delta G = 0.8$ or $1.9$ $k_BT$ (Figure 4b), respectively, which is accessible via thermal diffusion and can be enhanced by the applied electrical field.

To explain the contrast between the conductivity of ions, we calculated the hydration radii $R_H$ of the ions and conclude with the order K$^+$ < Na$^+$ < Li$^+$ < Ca$^{2+}$ < Mg$^{2+}$ (Figure 4c), which indicates that ion translocation measured in our experiments is manifested in a size-sieving mechanism. This aligns with the fact that the thickness of water layer is comparable with $R_H$ plus the van der Waals distances. These results also suggest that one could further engineer the functional groups of graphene edges to gain control of the selectivity[34].

The existence of a finite free energy barrier $\Delta G$ on the order of $k_BT$ indicates a prominent gating effect on the ion translocation process. To explore the gating effect, we carry out non-equilibrium MD simulations by applying an external field $E_y = 0.01\text{-}1.00$ eV/nm, and counting the probability of transmission and rejection events (Figure 4d). The results suggest that the transmission probability of ions through the water film measured in a fixed time interval increases with the field strength, demonstrating less torturous trajectories.

**Conclusions**

In summary, graphene nanopore devices occluded by a surface nanobubble demonstrated strong inter-cation selectivity, and ionic transport was modulated by an applied gate voltage. By utilizing pores in a de-wetted state, we have demonstrated a cation selective solid-state nanopore device that does not rely on the controlled fabrication of sub-nm pores. MD simulation results indicate that the ion selectivity can be explained by ion transport occurring across thin water films along the edge of the graphene pore, with transmission across the pore highly dependent on an externally applied electric field. Development of a defined process for control of nanobubbles will be necessary for further enhancing selectivity control. This ability to control selective nanopores at low voltages (< 500 mV) and with biologically relevant concentrations (100 mM) is an exciting advancement in sensing and separation technologies, not only providing a solid-state analog to voltage-gated biological ion channels, but having potential for applications in nanofluidic circuitry, water filtration, and energy storage as well.

**Conflicts of interest**

There are no conflicts to declare.

**Author Contributions**

L.C. and D.L. performed the experiments. L.C., J.L.S. and J.S.B conceived and designed the experiments. L.C., J.L.S., S.H., P.V.B., D.L. prepared and fabricated the samples. J.L.S., D.A.C., and A.J.R. performed HIM and STEM perforation and imaging. D.L. performed AFM imaging. K.Z. and Z.X. carried out molecular dynamics simulations. L.C., J.L.S., D.L., D.A.C., A.J.R., P.V.B., K.Z, Z.X., S.S. and J.S.B. interpreted the results and co-wrote the manuscript.

**Acknowledgements**

The authors acknowledge K. Ekinci for the use of the optical table, C. Duan for lab use and useful discussions. This work was funded by the National Science Foundation (NSF), grant no. 1054406 (CMMI: CAREER, Atomic Scale Defect Engineering in Graphene Membranes), grant no. 1706322 (CBET: Bioengineering of Channelrhodopsins for Neurophotonic and Nanophotonic Applications) and by the NSF Graduate Research Fellowship Program under grant no. DGE-1247312. HIM perforation and aberration-corrected STEM imaging were conducted at Oak Ridge National Laboratory's Center for Nanophase Materials Sciences (CNMS), a U.S. Department of Energy Office of Science User Facility.

**Notes and references**

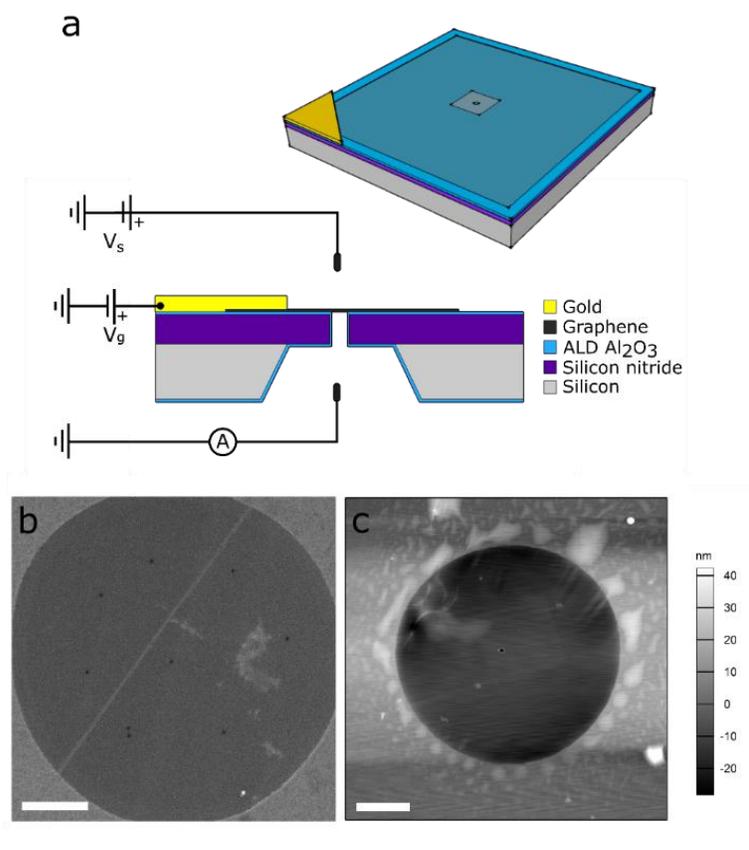

**Figure 1. Experimental set up.** (a) 3D rendering of the device along with schematic of the measurement circuit and cross section of the device. Graphene is suspended over a 5 µm hole in the silicon nitride window and mounted in a custom microfluidic cell in which electrolyte solution is introduced to both sides of the graphene membrane. (b) HIM image of CVD graphene with nine 35 nm pores drilled using HIM. (c) AFM image in air of CVD graphene with a single 50 nm pore drilled using HIM. Scale bars = 1 µm.

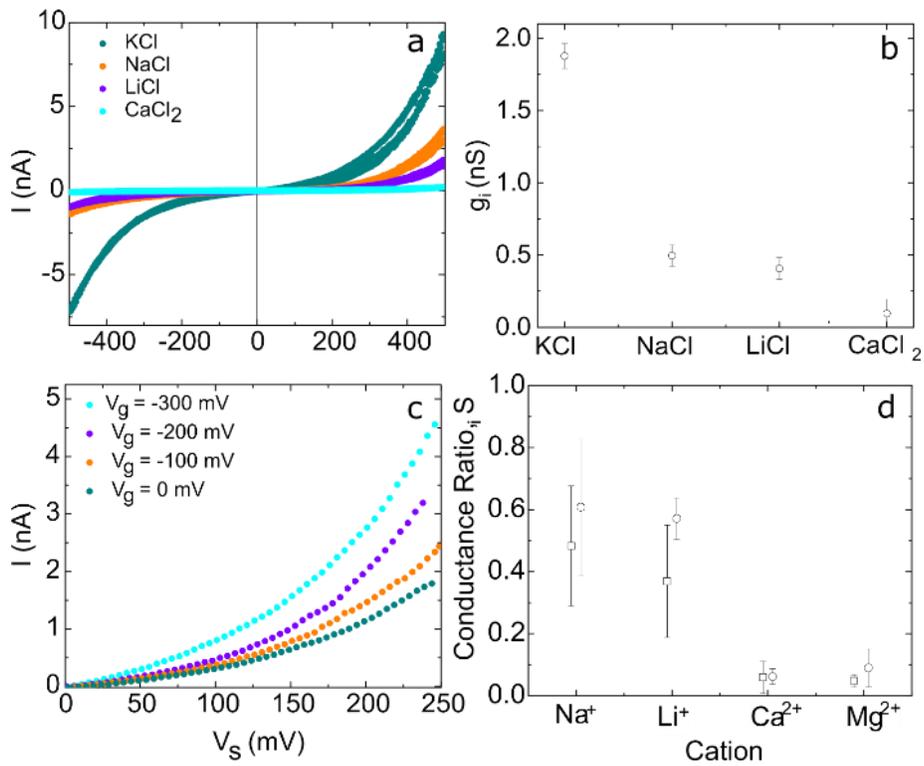

**Figure 2. Current-voltage characteristics.** (a) I-V curves and (c) gating behavior for the device in figure 1b. (b) Conductance taken at $V_s = 0$ mV for various electrolyte solutions. X-axis is ordered from lowest to highest cation hydration energy. All solutions are at 0.1 M chloride concentrations. (d) Inter-cation conductance ratio ($S_i$) of graphene nanopores sorted by cation. Open squares and open circles represent mean and standard deviation for devices with nine ~30 nm pores (for $S_{Na}$: N=5, $S_{Li}$, $S_{Ca}$: N=4, $S_{Mg}$: N=3) and devices with a single ~50 nm pore ($S_{Na}$, $S_{Li}$, $S_{Mg}$: N=3, $S_{Ca}$: N=2), respectively.

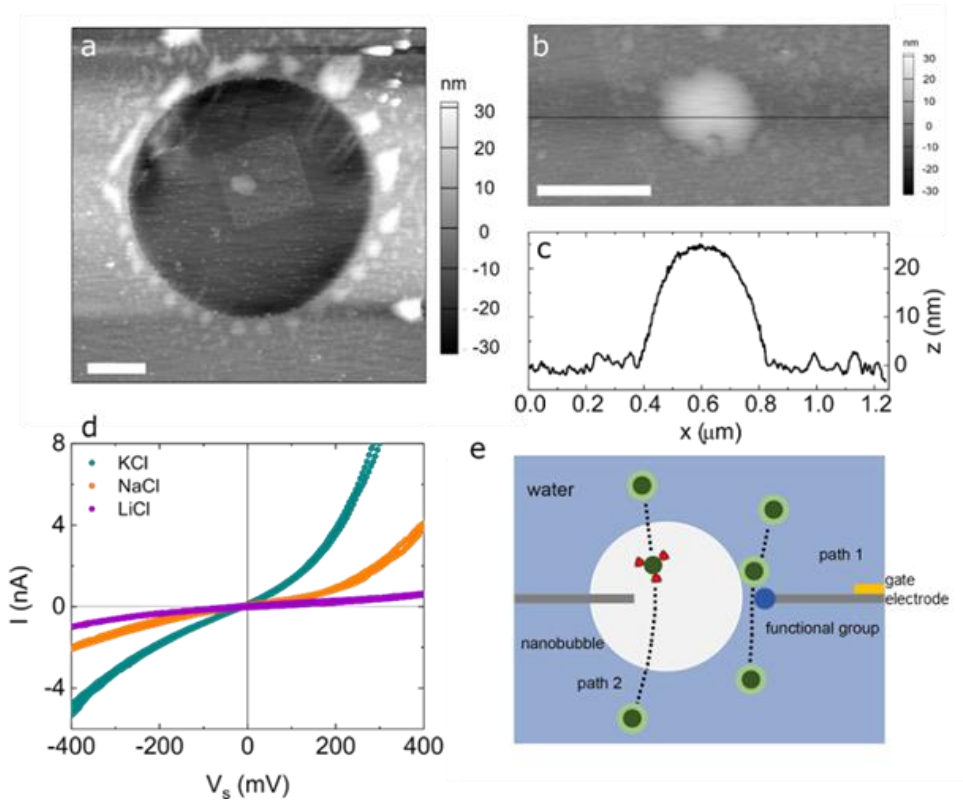

**Figure 3. Liquid AFM imaging.** (a) AFM image of a graphene membrane in water reveals bubbles on the surface of the graphene (compare to Fig. 1c of the same device measured in air). The square outline surrounding the large bubble corresponds with the area of graphene exposed during HIM drilling, which likely modified the surface of the graphene. Scale bar = 1 μm. (b) High magnification AFM image of bubble over pore (scale bar = 500 nm) and (c) corresponding cross sectional line cut. (d) Conductance measurements across device shown in (a) demonstrate non-linear, selective I-V behavior. (e) Schematic illustration of the two transport pathways considered during MD simulations. Gray planes represent the graphene and the dark blue circle on the edge is functional group. The dark green circles are ions and the outer light green parts are hydration water. The inner white region indicates the bubble. In path 1, the ion travels along the edge of the graphene nanopore, whereas in path 2, ions are transported through the water/gas interface.

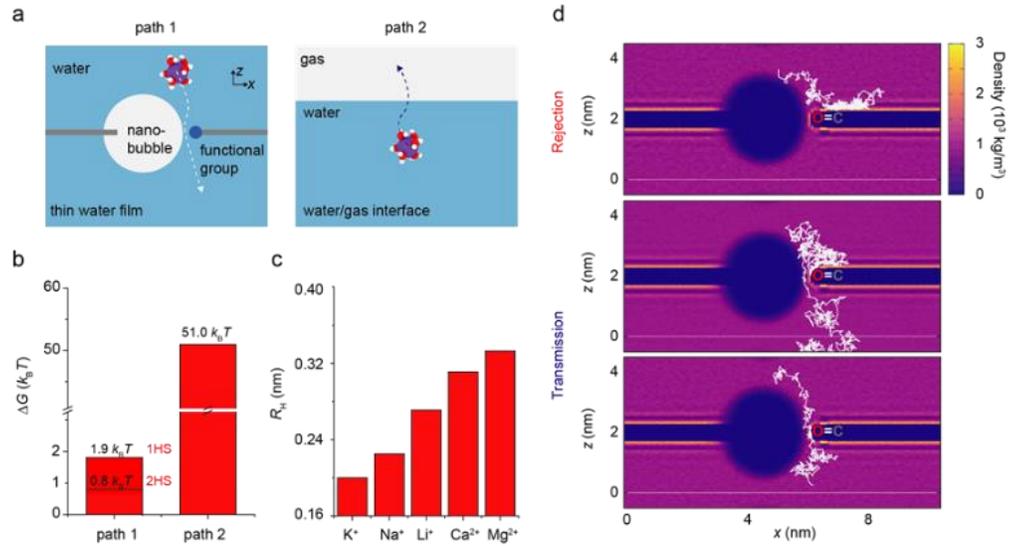

**Figure 4. MD simulations.** (a) Illustration of molecular simulation models and transport pathways (annotated by the dash arrows). Along path 1, ions travel along the edge of graphene nanopore coated by a thin water film, experiencing a free energy barrier of a few $k_BT$. Along path 2, ions are transported through the water/gas interface, experiencing a remarkably high free energy barrier of 51 $k_BT$. (b) Free energy barriers calculated for path 1 and 2, where 1HS/2HS indicate the situation where the 1st/2nd HS is perturbed by the functionalized graphene edge. (c) Hydration radii of ions for ions confined in nanometer-thick water films. (d) Ion diffusion under an external electric field. The ions can be translocated or rejected, demonstrating the gating effect through a thin water layer coating graphene edges, which can be controlled by the strength of external electric field. Color indicates the density of water. The atomic trajectories are extracted from simulations with an electric field applied in the z direction. The strength is 0.01, 0.1, and 1.00 eV nm$^{-1}$, for the results, from top to bottom respectively.

# Voltage gated inter-cation selective ion channels from graphene nanopores

*Lauren Cantley[1], Jacob L. Swett[2], David Lloyd[1], David A. Cullen[3], Ke Zhou[5], Peter V. Bedworth[2], Scott Heise[2], Adam J. Rondinone[4], Zhiping Xu[5], Steve Sinton[2], J. Scott Bunch[1,6]\**

**Methods:**

**Device Fabrication.** 200 nm thick low-stress silicon nitride windows were patterned using standard photolithography followed by an anisotropic KOH etch of the underlying silicon. A 3-5 µm diameter through hole was patterned in the center of the silicon nitride window via reactive ion etching (RIE) using an $SF_6$-Ar plasma. 20 nm of ALD $Al_2O_3$ was deposited, conformally coating the support chip. After a pre-synthesis anneal of the Cu foil substrate, CVD graphene was grown via low-pressure CVD. Following synthesis, the graphene, while still on the Cu foil substrate, was irradiated with 500 eV $Xe^+$ ions for a total fluence of $3.7 \times 10^{13}$ $Xe^+/cm^2$. This process has been shown to not introduce visible >nm hole defects when analyzed by STEM. The Cu foil substrate was then etched away with ammonium persulfate and the CVD graphene was transferred, without a sacrificial carrier layer, over the 5 µm through hole in the silicon nitride window. Once transferred, the graphene was exposed to 180 seconds of UVO to reduce hydrocarbon surface contamination. Through STEM imaging, this was confirmed not to introduce defects into the single-layer graphene itself. The gate electrode was defined via shadow mask lithography to be in contact with the graphene and 5 nm chromium followed by 60 nm gold were deposited via thermal evaporation.

CVD graphene samples were perforated via site specific focused helium ion drilling. For HIM drilling, pores were drilled using a Zeiss ORION NanoFab scanning HIM equipped with a gas field ion source (GFIS) using He$^+$ ions at 25 kV. Small pores were drilled with a stationary beam in defined locations with a beam current of 0.5-2.0 pA, and a dwell time of 52 ms per pore. Larger pores were drilled by defining a pattern and dwelling the beam with a 1 nm pixel spacing over that region with a dose on the order of 1 nC/µm$^2$ (1x10$^{18}$ He$^+$/cm$^2$).

**I-V and conductance measurements.** Conductance measurements were carried out in a custom made microfluidic cell fabricated via standard PDMS soft lithography. The structure allowed for electrolyte solution to be introduced to both sides of the membrane, while still allowing for AFM access to the top side of the membrane. Two microchannels, 50 µm tall and 2000 µm wide, were patterned into PDMS and bonded to a glass slide. Four inlet ports are punched into the PDMS, with the graphene on silicon nitride chip mounted over the center ports that connect the two underlying microchannels. PDMS was painted over the edge of the chip to create a seal and isolate volumes on either side of the chip. The device was dried for 2 days at room temperature to allow the PDMS seal to cure.

Graphene membranes were first rinsed with ethanol for 2 minutes followed by deionized water to facilitate wetting of the nitride membrane. Once graphene devices were introduced to fluid, membranes remained in solution for the duration of the experiments. Devices were exposed to electrolyte solution for 5-10 minutes before conductance measurements were carried out. Solutions were left unbuffered to avoid any potential interactions between the graphene and solutes. For the duration of the experiments, all solution pHs ranged from 5.91 – 6.38. To change the electrolyte solution, the microfluidic cell was first flushed with deionized water for 8 minutes, and then the new electrolyte was introduced. To ensure the observed selectivity was not an artifact of switching between solutions, the order in which different salt solutions were measured was varied, and we repeatedly alternated back and forth between salt solutions to confirm the differences in conductance were stable.

If a device did not display appreciable conductance upon initial wetting, it was exposed to ethanol for additional time. The I-V curves used for selectivity measurements within the text correspond to when a device was first witnessed to have appreciable conductance in aqueous solution. For extended storage, devices were stored in either deionized water or 0.1M KCl at 4°C.

I-V characteristics were measured with a DL Instruments model 1211 current preamplifier with silver/silver chloride electrodes at a sampling speed ranging from 1 – 10 mV/s. The voltage was ramped up and back down for each I-V measurement.

In order to measure selectivity, the concentrations used to measure conductance were chosen to maintain a constant chloride ion concentration across experiments comparing both monovalent and divalent ions; for example a 100 mM concentration for monovalent ions and a 50 mM concentration for divalent ions. This ensured that differences in conductance were due to differences in ion mobility and not from differences in bulk conductivity. Conductance values in the main text were determined by taking a linear fit of the IV curve for +/- 50 mV around $V_s = 0$ mV, unless otherwise noted.

Bulk conductivities of the solutions used in this study were measured using Mettler Toledo S230 SevenCompact™ conductivity meter with InLab® 731 ISM probe for concentrations $\geq 10^{-2}$ M and InLab® 741 ISM probe for concentrations $< 10^{-2}$ M.

**Imaging.** Aberration-corrected STEM imaged graphene devices were imaged in a Nion UltraSTEM 100 operated at 60 kV. Prior to imaging, the devices were baked at 160°C for 6-8 hours in a vacuum chamber to remove mobile hydrocarbon contamination. Aberration-corrected STEM images were acquired using a medium-angle annular dark-field (MAADF) detector with a 54-200 mrad collection semi-angle. A convergence semi-angle of 30 mrad and probe current of 60-80 pA was used to acquire atomic resolution images of the pore edges.

HIM imaged graphene devices were imaged in a Zeiss ORION NanoFab operated at 25 kV with $He^+$. No special preparation was performed prior to drilling and imaging. To limit the ion dose to the samples, images were acquired at a large field of view relative to the pores (typically 5 x 5 µm), with beam currents between 0.5-2.0 pA, 512x512 resolution, and 2-50 µs dwell time per pixel.

AFM images were taken with an Asylum MFP-3D microscope operated in tapping mode. Platinum-iridium coated silicon AFM tips (ARROW-NCPt) were used for air imaging. For fluid imaging, $Si_3N_4$ tips (PNP-TR, spring constant $k_c$ = 0.32 N/m) were driven at a free air amplitude of around 50 nm. After tip engagement, the set point amplitude was adjusted to be as high as possible, set just below the value at which the tip lost contact with the surface. This meant the tip would exert the smallest possible forces and avoid modifying the topology of the sample being measuring. Scanning speeds were approximately 3 µm/s, and post-processing of the images was carried out to remove low frequency noise.

**Molecular Dynamics Simulations.** We perform MD simulations using the large-scale atomic/molecular massively parallel simulator (LAMMPS)[1]. A quasi-two-dimensional pore-bubble model is constructed with a thickness of 2.91 nm along the *z* direction (Figure 3e) to minimize the size effect in the MD simulations[2]. The all-atom optimized potentials for liquid simulations[3] are used for the graphene. The SPC/E model is used for water, which is widely adopted for MD simulations of water transport as it predicts reasonable static and dynamic propensities[4]. The van der Waals interactions between water and the nanoporous graphene membrane are modeled in the Lennard-Jones (L-J) 12-6 form, *i.e.* $V = 4\varepsilon[(\sigma/r)^{12} - (\sigma/r)^6]$. The zigzag edge of graphene is terminated by hydrophilic carbonyl groups with a density of one group per under-coordinated carbon atoms, following the experimental evidence[5]. The interaction between carbon atoms in graphene or GO and oxygen atoms in water is modeled with parameters $\varepsilon_{C-O}$ = 4.063 meV and $\sigma_{C-O}$ = 0.319 nm. This set of parameters predicts a water contact angle of 98.4° for graphene, consistent with the value measured experimentally[6]. For the ions, we use the force-field parameters developed by Kenneth's group, listed in Table S1, which are optimized for a reliable description of the free energy and shell structure of ion solvation, with long-range electrostatic interactions treated by using the particle-mesh-Ewald (PME) method[3]. A Weeks-Chandler-Andersen (WCA) gas model[4] is used to simulate the bubble, with interatomic interaction described using a 12-6 L-J potentials ($\varepsilon$ = 0.0103 eV, $\sigma$ = 0.34 nm), truncated and shifted at interatomic distance $r = 2^{1/6}\sigma$. Hetero-atomic parameters for the 12-6 L-J interactions are determined through the Lorentz-Berthelot mixing rules.

The translocation pathway and PMF of ion transport along paths **1** and **2** (Figure 4a) are determined by umbrella sampling (US)[5]. We use a harmonic-spring biasing to constrain the ionic position at a specific value of $d$ (Figure S1a). For path **1** in the film and the fully-immersed model models (model 1 and 2 in Figure S1a), the force constant is set to $10^4$ and $10^3$ kJ/mol/nm$^2$ for $d$ = 0.20-0.35 nm (0.05 nm per window) and $d$ = 0.4-1.0 nm (0.10 nm per window), respectively. For path **2**, a force constant of $2\times10^3$ kJ/mol/nm$^2$ is used (0.10 nm per window). Every window runs in a separate simulation for 6 ns and the trajectories are collected every 250 ps. The distribution of atomic positions in adjacent windows significantly overlaps. Finally, the PMF is generated by recombining individually-biased distributions using the weighted histogram analysis method[5], by using the package implementing by Grossfield[7]. The PMF profiles for path 1 (models 1 and 2) and path 2, using Na$^+$ as an example without the loss of generality, are shown in Figure S1b. For model 1 and model 2, the two distinct valleys correspond to the condition that the 1$^{st}$/2$^{nd}$ HS is perturbed by the functionalized graphene edge (Figure S1c). The free energy barrier for translocation in model 2 is 2.0/1.0$k_B T$, respectively, close to the values measured in model 1 (1.9/0.8 $k_B T$).

The hydration radii of ions confined within thin water films are calculated following our previous work[8], by using a model with a 1-2.5 nm-thick water film on graphene (model 3 in Figure S1).

**Control Measurements.** Several control experiments were carried out to verify that the measured current was passing through the graphene pore(s) and not leaking through the graphene/silicon nitride interface or elsewhere. The following control devices were measured:

(1) ALD alumina coated silicon nitride membrane with no through-hole, no graphene. Isolate transport through silicon nitride membrane and through the PDMS sealing the device within microfluidic cell.

(2) ALD alumina coated silicon nitride membrane with 5 µm through-hole, no graphene. Determine limiting conductance for nitride through-hole.

(3) ALD alumina coated silicon nitride membrane with 5 µm through-hole, suspended unperforated CVD graphene.

Leakage conductance was measured to be less than 55 pS in 0.1 M KCl and less than 140 pS in 1.0 M KCl for type (1) control devices. The limiting conductance through type (2) control devices was measured to be on the order of µS for 0.1M KCl. For type (3) control devices, the conductance was less than 60 pS in 0.1M KCl and less than 280 pS in 1.0 M KCl.

**Asymmetric ion conditions.** In the main text, we concluded that cations were the majority charge carriers due to the selective I-V behavior and p-type gating behavior. However, there is also the distinct possibility that the current is carried by anions, but modulated by cations. To address this we carried out conductance measurements with asymmetric ion conditions: 0.05 M $CaCl_2$ on one side of the membrane and 0.1 M KCl on the other. Concentrations were chosen to maintain a constant chloride concentration. In this configuration, we observed highly rectified behavior, consistent with the conclusion of cations as the majority charge carriers (Fig. S2).

**Pore Termination.** High magnification aberration-corrected STEM imaging was carried out to reveal the pore geometry, edge structure, and the surface of the graphene (Fig. S3). Carbon was the most prevalent edge termination in the pores imaged, however hydrocarbon contamination, which is prevalent across the surface of the graphene, was also present near the pore edges for several of the devices that were imaged.

**Evidence for the presence of a bubble.** We found further evidence for the presence of a bubble occluding the pore by varying tip-sample forces during scanning[9]. This was done by reducing the amplitude set point ratio (the ratio between the amplitude set point and the free air amplitude), which increases the forces on the sample from the tip. Figure S4 shows the result of imaging the pore in water with different set points ratios. The lower set point ratio of 30% caused the tip to deform the object over the pore (Fig. S4B) as compared with images using higher set point ratios of 36% (Fig. S4A). This means the object is soft and malleable, and has a similar response to AFM probing as seen in PDMS nanodroplets or gaseous nanobubbles[10]. During the subsequent scan the set point was raised again causing the tip-sample forces to reduce (Fig. S4C), and the bubble topology returned to the same state as before.

**Conductance measurements in solvents of varying surface tensions.** We performed liquid AFM imaging on the device in two solvents with different surface tension: water (72.86 mJ/m$^2$)[11] and ethanol (22.39 mJ/m$^2$)[12]. For the device shown in figure 1c, liquid AFM imaging revealed that the surface nanobubble occluding the pore in aqueous solution (Fig. S5A) was absent in ethanol (Fig. S5B), and returned when the solvent was exchanged back to water (Fig. S5C). The device was not dried out and remained continuously wet between solvent exchanges.

Additionally, Li$^+$ transport across three devices was measured in both aqueous and ethanol based solutions. Bulk conductivities were measured to be 10.7 mS/cm for aqueous 0.1 M LiCl and 1.54 mS/cm for the ethanol based 0.1 M LiCl. Measurements are carried out in the order shown in the legend, seamless switching between ethanol and aqueous based LiCl solutions to keep the device wet at all times. Devices demonstrate increased conductance across the pore for ethanol based solution when compared to aqueous solution of the same ionic concentration. (Fig. S6A-C). The bulk conductivity of 0.1M LiCl in ethanol is considerably less than 0.1M LiCl in H$_2$0[11,12]. Thus, one would not expect the conductance across the graphene pore to be greater in the ethanol solution unless the transport pathway changed between the two measurements. This change in transport pathway is due to the presence and absence of a nanobubble, witnessed via liquid AFM imaging (Figure S4).

In situ measurements were also carried out to allow for monitoring of the conductance while switching between ethanol and water solutions (Figure S6D). Sweeping the bias voltage, the device was first measured in 0.1M LiCl in $H_2O$ (blue arrow) and demonstrated little conductance. As 0.1M LiCl in ethanol was introduced and the aqueous solution flushed out, the conductance increased (orange arrow) and stayed at the elevated conductance level while the ethanol solution was present (green arrow). All the while, the voltage was continually swept, down to -500 mV and back to +500 mV. When the water solution was reintroduced and the ethanol solution flushed out, the conductance returned to the lower conductance state (purple arrow) and remained low while water solution was present (yellow arrow). This result was consistent with measurements in figure S6A-C.

**AFM and conductivity measurements of a non-occluded pore.** We performed conductivity and in-situ AFM measurements of nanopores which were not occluded. Figure S7A shows an AFM image of a HIM drilled nanopore imaged in water. The phase channel was used to plot the image in order to highlight regions of rapid change in the topology of the surface – in this case the edge of the nanopore (Fig. S7C). Immediately prior to AFM imaging, I-V curves of LiCl, NaCl and $CaCl_2$ solutions all demonstrated linear conductance (Fig. S8). We can estimate the nanopore diameter, $d$, which should give rise to a conductance $G_i$ for each salt species $i$ by using the theory for ion transport through a cylindrical pore, which yields the formula:

$$G_i = \sigma_i d \tag{S1}$$

where $\sigma_i$ is the bulk conductivity, and access resistance is assumed to be the dominant term[13]. Using values for $G_i$ calculated from Fig. S7 and values for $\sigma_i$ measured prior conductance measurements, Eq. S1 was used to calculate the predicted pore diameter $d$ for each salt and tabulate the results in Table S2.

The pore conductivities for three different salts give consistent predictions for a pore size of ~75 nm in diameter. Comparing this prediction with the pore size measured with AFM in Fig. S7B, we see there is close agreement between the two. This confirms that when no occlusion is present over the pore, the observed nanopore conductance is consistent with the behavior we would expect for an infinitely thin cylindrical pore of that diameter.

**Prevalence and formation of nanobubbles.** To estimate the prevalence of nanobubbles in our experiments, we imaged six devices in water with AFM. We found the presence of nanobubbles in all six devices, four of which are shown in Fig. S10 as an example. Nanobubbles are distributed across the devices and are of various sizes. We found evidence of bubbles occluding the graphene nanopore in four of the six devices measured. The ethanol and water solutions were degassed in a vacuum desiccator for ~1h prior to measurement in three of these devices. In addition, we found that pores frequently had a conductance lower than is predicted analytically, given the imaged diameter of the pore. We interpret this as indirect evidence that the pore was obstructed due to bubble formation. We measured the pore diameter and conductance of aqueous salt solutions across 35 devices, including the nine devices highlighted in the main text, only two devices displayed a conductance magnitude consistent with what is predicted analytically for the imaged pore size. The other 33 devices all displayed lower than expected conductance values.

We hypothesize that the presence of hydrocarbon contamination around the pore mouth makes the pore region more hydrophobic, and therefore more prone to dewetting. This hydrocarbon contamination can be seen in Fig S11 which shows a HIM image of a device after pore milling. The bright regions around the pore are caused by charging of the electrically insulating hydrocarbons stuck to the graphene surface.

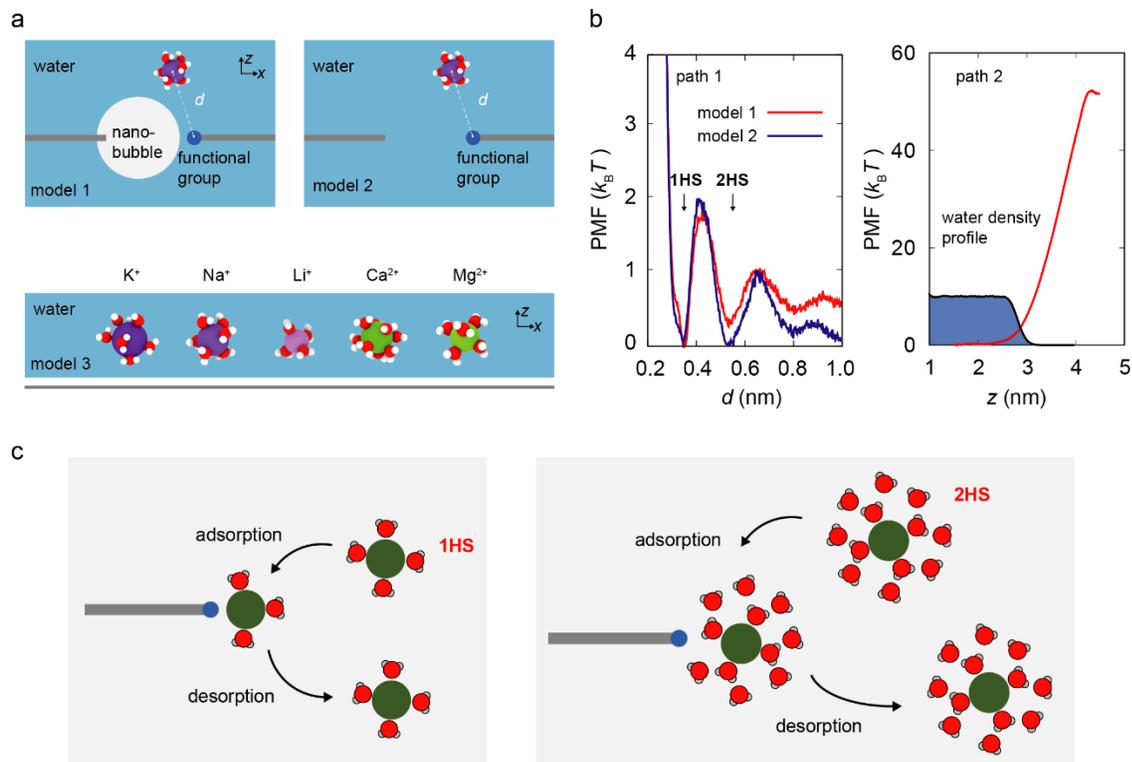

**Figure S1**. (a) Illustration of molecular simulation models: Models 1 and 2 for the PMF of ions translocated with hydration shells perturbed by the functionalized graphene edges, which are coated by a thin water film or immersed in water, respectively. Model 3 for the calculation of hydration radii of ions. (b) PMF profiles for path 1 (models 1 and 2) and path 2. (c) Illustration of the adsorption-desorption process of ion translocation through the thin water layer, with the 1st or 2nd hydration shells (1HS/2HS) perturbed by the functional groups.

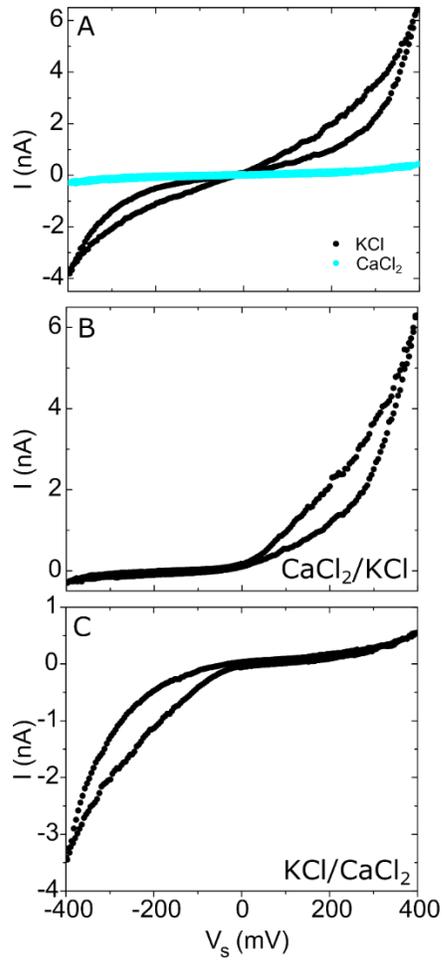

**Fig. S2. Conductance under asymmetric ion conditions.** (A) Symmetric ion conditions for both KCl and $CaCl_2$ at 0.1M chloride concentration. (B), (C) Asymmetric ion conditions in 0.1M chloride concentration.

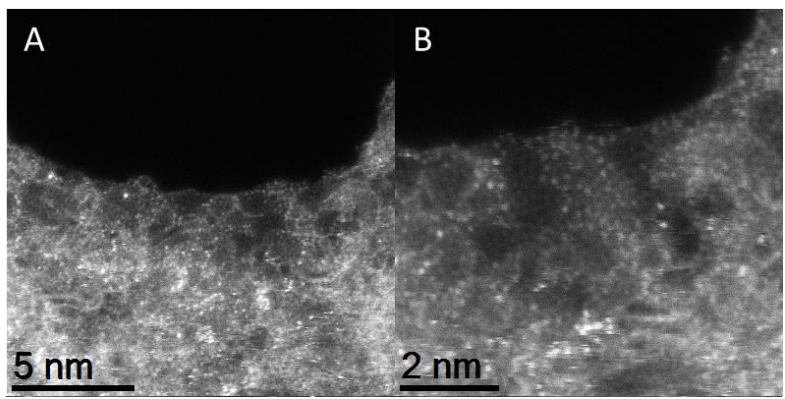

**Fig. S3. Pore termination.** Aberration-corrected medium-angle annular dark-field (MAADF) STEM images of CVD graphene (A) HIM drilled pore periphery, (B) HIM drilled pore edge showing nanopore geometry, edge structure, and hydrocarbon contamination.

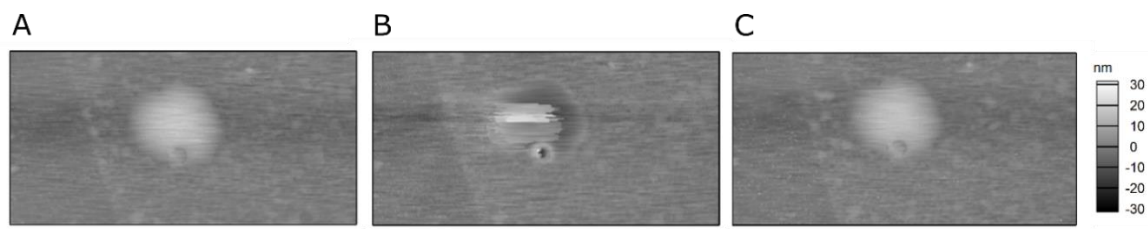

**Figure S4.** AFM images of the pore taken in water with different set point ratios of (A) 36%, (B) 30%, and (C) 36% again. The lower set point ratio increases tip-sample forces, which deform the bubble.

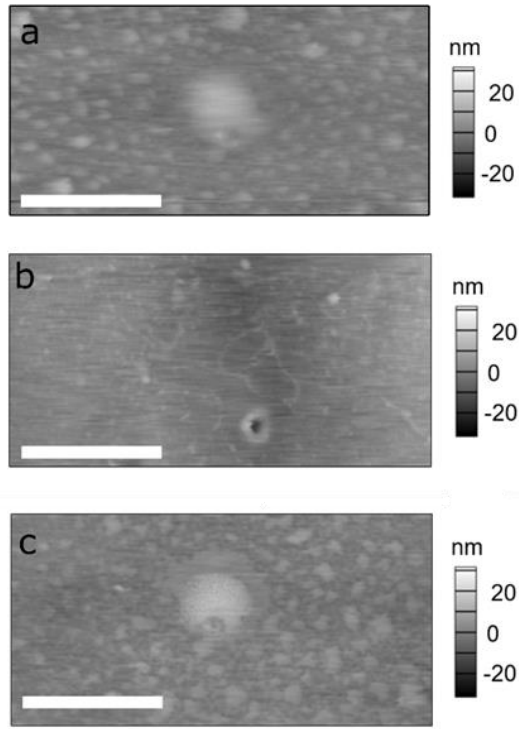

**Figure S5. Pore Wettability.** An ethanol/water exchange was monitored via liquid AFM imaging: (A) graphene pore in water with a surface nanobubble, (B) pore in ethanol solution with no surface nanobubble, and (C) the nanobubble returns when solution is exchanged back to water. Scale bars = 500 nm

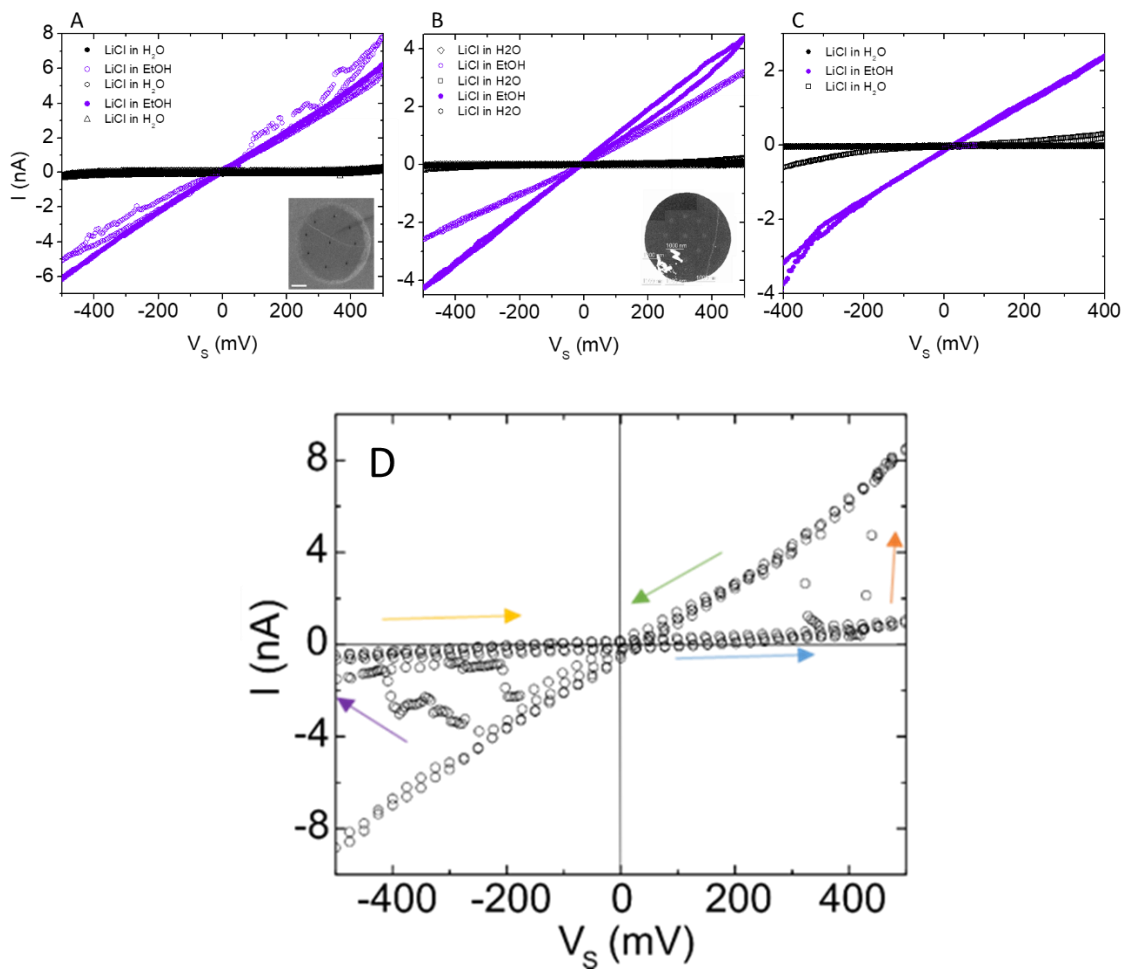

**Fig. S6. Conductance in solutions of different surface tension.** . (A-C) Aqueous and ethanol based 0.1M LiCl conductance across graphene nanopore device. Inset depicts (A)HIM and (B) STEM image of the device measured. Scale bar = 1 micron. Measurements are carried out in the order shown in the legend. The conductance of the LiCl in ethanol was greater than that of LiCl in water. (D) In situ measurement of conductance while switching between water and ethanol based 0.1M LiCl solutions.

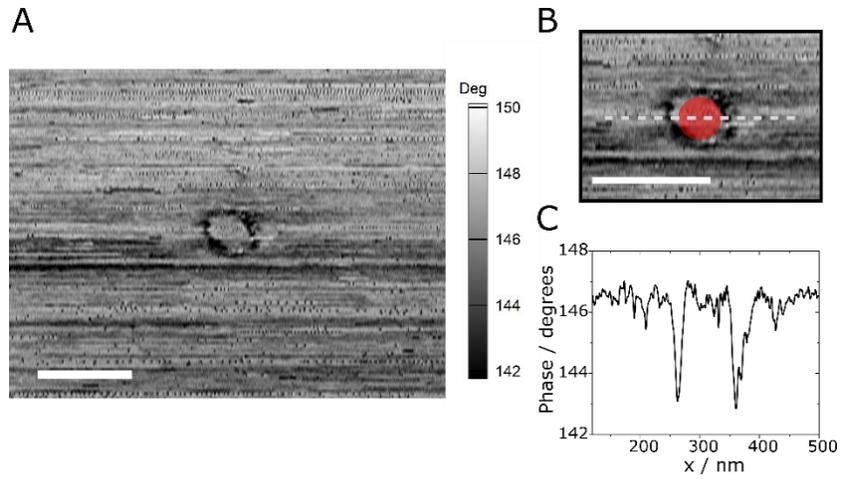

**Figure S7. AFM imaging of non-occluded nanopore.** A) AFM phase channel image of a HIM drilled nanopore. B) Expanded view of nanopore, with a comparison to the 75 nm circular pore implied by our conductance data. C) Phase line-cut across the pore (as shown in B) showing the outline of the pore. Scale bars are 200 nm.

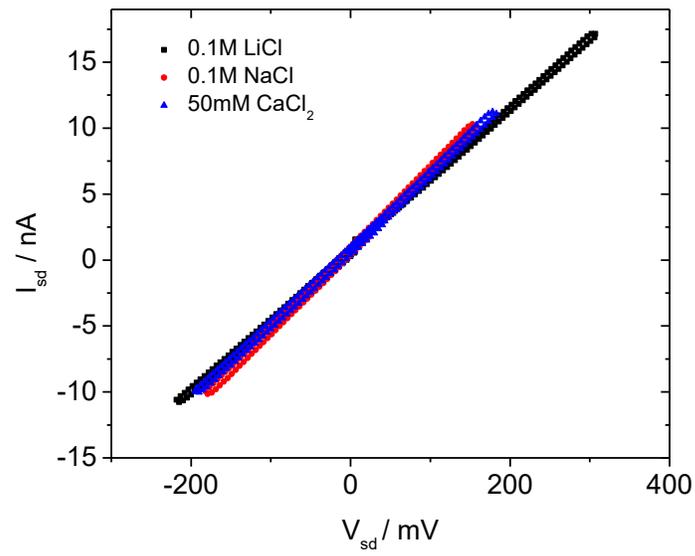

**Figure S8**. Two terminal I-V curves of the pore for different salts, taken immediately prior to the AFM measurements.

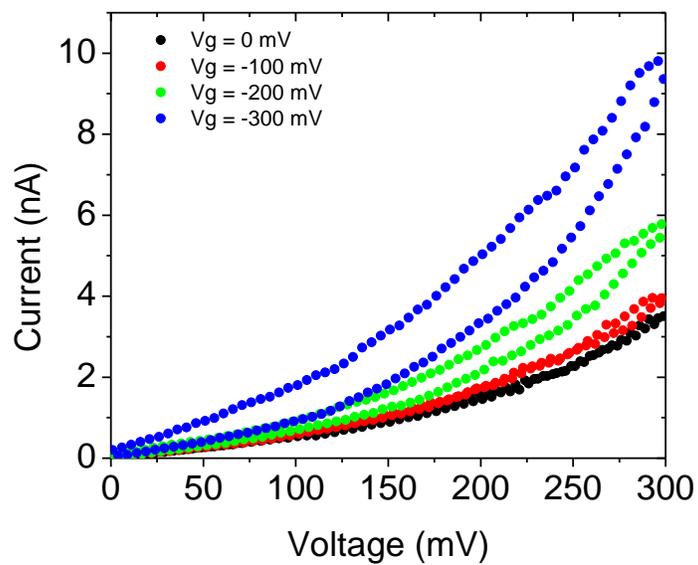

**Figure S9.** Typical hysteresis observed from applied gate voltage while sweeping source/drain voltage from 0 mV to 300 mV and back down to 0mV, in a 0.1 M KCl concentration solution.

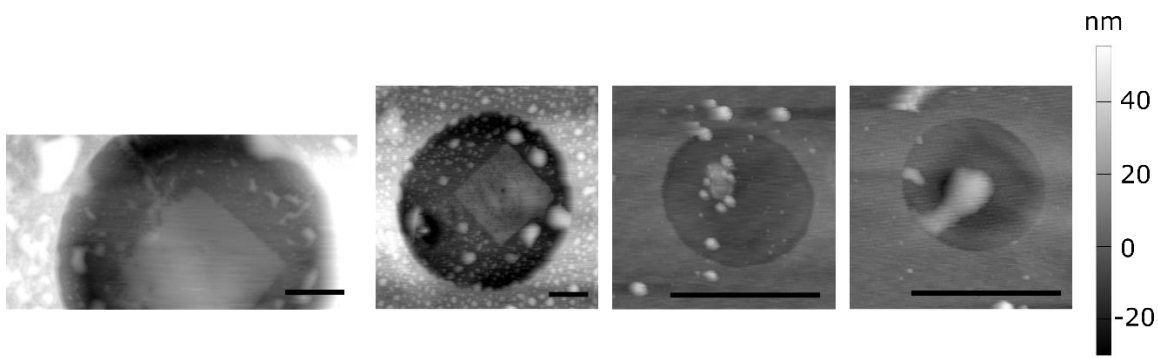

**Figure S10.** Examples of other devices we imaged which are covered in nanobubbles. The devices were measured with AFM in water. Scale bars are 1μm.

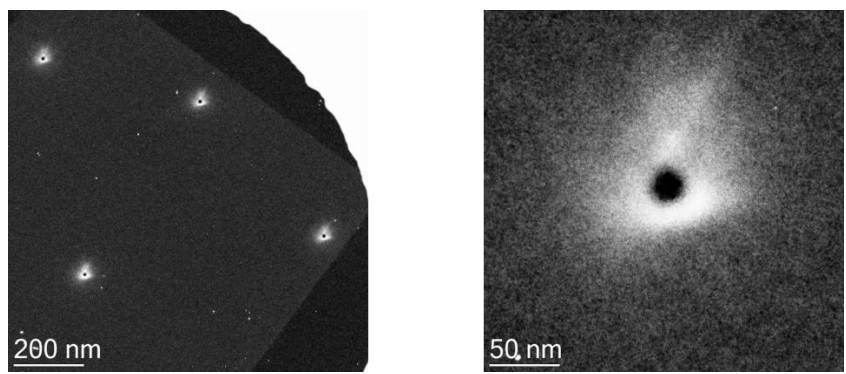

**Figure S11.** Evidence of hydrocarbon contamination around the nanopore mouth after HIM milling. The white region indicates charging is occurring on the electrically insulating hydrocarbon layer.

| Ions | $\varepsilon$ (kcal/mol) | $\sigma$ (nm) |
|---|---|---|
| $Na^+$ | 0.02639 | 0.25907 |
| $K^+$ | 0.12693 | 0.29988 |
| $Li^+$ | 0.00274 | 0.22415 |
| $Ca^{2+}$ | 0.09788 | 0.29132 |
| $Mg^{2+}$ | 0.01020 | 0.24232 |

**Tables S1.** 12-6 Lennard-Jones (L-J) potential parameters used for the ions studied[6].

| Salt solution | Bulk Conductivity (S/m) | Pore Conductance (nS) | Estimated pore size (nm) |
|---|---|---|---|
| 100 mM LiCl | 0.814 | 59.4 | 73.0 |
| 100 mM NaCl | 0.884 | 65.7 | 74.3 |
| 50mM CaCl$_2$ | 0.834 | 62.3 | 74.7 |

**Table S2.** Bulk conductivities for each salt solution were measured prior to our experiments. The pore conductance was calculated from the data in Fig. S7, and was used with Eq. S1 to estimate the pore size.